\documentclass[pdftex,twocolumn,epjc3]{svjour3}          

\RequirePackage[T1]{fontenc}

\smartqed  

\RequirePackage{graphicx}
\usepackage{latexsym}
\usepackage{multirow}
\usepackage{amsmath,amssymb,amsfonts,latexsym,cancel}
\usepackage{subcaption}
\RequirePackage{mathptmx}      
\RequirePackage{flushend}
\RequirePackage[numbers,sort&compress]{natbib}
\RequirePackage[colorlinks,citecolor=blue,urlcolor=blue,linkcolor=blue]{hyperref}

\journalname{Eur. Phys. J. C}

\begin{document}

\title{Gravitational wave asteroseismology of charged strange stars in the Cowling approximation: the fluid pulsation modes}

\author{Jos\'e D. V. Arba\~nil\thanksref{e1,addr1,addr2}
        \and
        C\'esar H. Lenzi\thanksref{addr3}
        \and
        Juan M. Z. Pretel\thanksref{addr4}  
        \and
        C\'esar O. V. Flores\thanksref{addr5,addr6}}
\thankstext{e1}{e-mail: jose.arbanil@upn.pe}

\institute{Departamento de Ciencias, Universidad Privada del Norte, Avenida el Sol 461 San Juan de Lurigancho, 15434 Lima, Peru \label{addr1} \and
Facultad de Ciencias F\'isicas, Universidad Nacional Mayor de San Marcos, Avenida Venezuela s/n Cercado de Lima, 15081 Lima,  Peru \label{addr2}
\and
Departamento de F\'isica, Instituto Tecnol\'ogico de Aeron\'autica, S\~ao Jos\'e dos Campos, SP, 12228-900, Brazil
\label{addr3}
\and
Centro Brasileiro de Pesquisas F{\'i}sicas, Rua Dr. Xavier Sigaud, 150 URCA, Rio de Janeiro CEP 22290-180, RJ, Brazil\label{addr4}
 \and
Universidade Estadual da Regi\~ao Tocantina do Maranh\~ao, UEMASUL, Centro de Ci\^encias Exatas, Naturais e Tecnol\'ogicas,  Imperatriz, CEP 65901-480, MA, Brazil\label{addr5}
\and
Universidade Federal do Maranh\~ao, UFMA, Departamento de F\'isica-CCET, Campus Universit\'ario do Bacanga, S\~ao Lu\'is, MA, CEP 65080-805, Brazil\label{addr6}}

\date{Received: date / Accepted: date}

\maketitle

\begin{abstract}
In this work we study, within the framework of Cowling approximation, the effect of the electric charge on the gravitational wave frequency of fluid oscillation modes of strange quark stars. For this purpose, the dense matter of the stellar fluid is described by the MIT bag model equation of state (EoS), while for the electric charge profile, we consider that the electric charge density is proportional to the energy density. We find that the gravitational wave frequencies change with the increment of electric charge; these effects are more noticeable at higher total mass values. We obtain that the $f$-mode is very sensitive to the change in the electric charge of the star. Furthermore, in the case of the $p_1$ mode, the effect of the electric charge is not very significant. Our results reveal that the study of the fundamental pulsation mode of an electrically charged compact star is very important in distinguishing whether compact stars could contain electric charge. We also employ another electric charge distribution profile that follows a power law, and it is found that the $f$-mode change is more noticeable than the $p_1$-mode when the electric charge is incremented.
\end{abstract}

\maketitle


\section{\label{sec:level1} Introduction}

With the recent detections of gravitational waves from the merger of compact binary systems, provided by the LIGO-VIRGO collaboration, a new golden era of general relativity (GR) has begun ~\cite{abbott2016,abbott2016_2,abbott2016_3,abbott2017,abbott2017_1,abbott2017_2,abbott2017_3,abbott2017_4,abbott2017_5,abbott_2018a_tidal}. These systems allow us to study the validity of GR in its strong-field regime and also make possible to investigate the behavior of matter under extreme conditions.

Under the conditions of binary coalescences, matter can be pushed to the region of high compactness and, consequently, hyper-massive neutron stars or black holes can be formed. If a neutron star is formed, it will accrete more matter from its environment, and, eventually, its internal structure can attain high densities. In those conditions, the production of strange quark matter is energetically favored. The\-refore, strange quark matter is a strong possibility and its study has been an object of very intense research \cite{bodmer1971,witten1984}. In this sense, if strange quark matter can be formed at high densities, two types of compact stars with that composition exist strange quark stars and hybrid stars. The first ones are composed of deconfined quarks and the second ones have a core of quark matter with an envelope of hadronic matter \cite{parisi2021}. Of course, there also exists the possibility of other exotic compositions for compact stars, for example, dark matter \cite{das_kumar2021,flores_lenzi_dutra2024}, dark energy \cite{2024PhRvD.109b3524P} and anisotropy \cite{arbanil_lenzi2023}.

We need more theoretical and astronomical tools to obtain more information about the dense matter contained with\-in compact stars. In that regard, future gravitational observations of the oscillation modes of compact stars can help to elucidate whether strange quark matter can be the true ground state of matter, as proposed by Bodmer \cite{bodmer1971} and Witten \cite{witten1984}. The gravitational wave asteroseismology of compact stars is therefore a very promissory tool that will allow us to understand the interior of compact stars through the investigation of their vibration modes.

The system of differential equations that govern those oscillations is obtained by replacing both the perturbed spacetime and the fluid variables into the Einstein field equation, into the conservation equation of the energy-momentum tensor, and into the conservation of baryon number; preserving only the first-order variables \cite{thorne1967,detweiler1983,detweiler1985}. The oscillation modes determined by using the complete linearized equations have been analyzed under different contexts; e.g., considering compact stars with a crust \cite{flores2017}, hadronic matter mixed with dark matter \cite{flores_lenzi_dutra2024}, and phase transition \cite{orsaria2019} (see also Ref.~\cite{tonetto_2020}).

A method widely used in the literature for the calculation of the oscillation mode frequencies (obtained from the complete nonradial oscillation equations) is the Cowling approximation \cite{mcdermott1983,lindblom1990}. It has been shown that such a method provides good precision results, with a discrepancy of less than $20\%$ for $f$-modes and $10\%$ for $p_1$-modes \cite{yoshida1997}. Based on these outcomes, this method is employed to investigate the pulsation modes under different frameworks. For example, this formalism has been applied to study nonradial oscillation frequencies of compact stars composed of a perfect fluid energy-momentum tensor \cite{vasquez2014,sotani2011,das_kumar2021}. There also exist works where rotation is considered \cite{stavridis2007,boutloukos2007}, crust elasticity \cite{samuelsson2007}, and anisotropic pressure \cite{doneva2012,arbanil_lenzi2023}, and even higher dimensions \cite{arbanil2020}.

In this work we investigate the nonradial oscillations of electrically charged stars within the Cowling approximation \cite{arbanil_malheiro2015}. For this objective, we investigate the influence of the electric charge on the frequency of the $f$- and $p_1$-modes of strange quark stars. For the dense matter, we employ the MIT bag model EoS and, for the charge fluid distribution, two profiles for the electric charge are considered; the charge density proportional to the energy density $\rho_e=\alpha\rho$ (with $\alpha$ being a dimensionless constant) and distribution of the electric charge proportional to the radial coordinate $r$ of the form $q=Q(r/R)^3$, with $Q$ and $R$ being the total charge and total radius. The present article is organized as follows: In section \ref{section2}, we present a general relativistic formulation for charged compact stars, where we derive the nonradial oscillation equations in the context of the Cowling approximation. Such equations now include extra terms corresponding to the electric charge and are hence the generalized version of previous studies. In section \ref{section3}, we investigate the charged strange stars by using the charge density profile $\rho_e=\alpha\rho$ and, in section \ref{section4}, we analyze the charged strange star by considering the electric charge function $q=Q(r/R)^3$. In these two last sections, we show and discuss the numerical results corresponding to the equilibrium configurations and the pulsation mode frequencies. Finally, in section \ref{section5}, we conclude. Along the article, the units $c=1=G$ are employed.

\smallskip

\section{General relativistic formulation}\label{section2}



\subsection{Einstein-Maxwell equations}\label{subsection1}

At first, we mention the main equations used in this article: the Einstein and the Maxwell field equations. Given a matter-energy distribution, the Einstein field equation used to deduce the spacetime metric is the following
\begin{equation}
 G_{\mu\nu}=8\pi T_{\mu\nu},\label{EFeq}
\end{equation}
and the Maxwell equations, used to determine the behaviour of the electromagnetic field in the curved spacetime, read as follows
\begin{equation}    
    \nabla_{\nu}F^{\mu\nu}=4\pi J^{\mu},\label{MFeq}
\end{equation}
with the Greek indices $\mu$, $\nu$, running from $0$ to $3$; $G_{\mu\nu}$ depicts the Einstein tensor and $T_{\mu\nu}$ stands for the energy-mo\-mentum tensor where we consider that $T_{\mu\nu}=M_{\mu\nu}+E_{\mu\nu}$. The tensor quantity $M_{\mu\nu}$ denotes the energy-momentum tensor for a perfect fluid
\begin{equation}\label{tem}
    M_{\mu\nu}=p g_{\mu\nu}+(\rho+p)u_{\mu}u_{\nu},
\end{equation}
where $\rho$, $p$, and $u_{\mu}$ are respectively the energy density, pressure, and the fluid's four-velocity. 

Moreover, $E_{\mu\nu}$ represents the electromagnetic energy-momentum tensor, namely
\begin{equation}\label{tem}
    4\pi E_{\mu\nu}=F_{\mu}^{\;\;\gamma}F_{\nu\gamma}-\frac{1}{4}g_{\mu\nu}F_{\gamma\beta}F^{\gamma\beta},
\end{equation}
with $F_{\mu\nu}$ representing the Maxwell tensor. In addition, the electric current density is given by
\begin{equation}\label{tem}
J^{\mu}=\rho_e u^{\mu},
\end{equation}
where $\rho_e$ is the electric charge density.

\subsection{Static stellar structure equations}\label{subsection2}

To describe a spherically charged fluid distribution, the metric is considered of the following form:
\begin{equation}\label{metric}
    ds^2=-e^{2\Phi}dt^2+e^{2\Lambda}dr^2+r^2d\theta^2+r^2\sin^2\theta d\phi^2,
\end{equation}
where $(t,r,\theta,\phi)$ represent the Schwarzschild-like coordinates, and with $\Phi=\Phi(r)$ and $\Lambda=\Lambda(r)$ being functions of the radial coordinate $r$ alone.

Taking into account the line element \eqref{metric} and metric function $\Lambda$ of the form: 
\begin{equation}\label{g00}
    e^{2\Lambda}=\left(1-\frac{2m}{r}+\frac{q^2}{r^2}\right)^{-1},
\end{equation}
the non-zero components of the Einstein-Maxwell equations \eqref{EFeq} and \eqref{MFeq} lead to
\begin{eqnarray}
&&\frac{dm}{dr}=4\pi\rho r^2+\frac{q}{r}\frac{dq}{dr},\label{mass_conservation} \label{eq_mass}\\
&&\frac{dq}{dr}=4\pi\rho_e r^2e^{\Lambda},\\
&&\frac{dp}{dr}=-(p+\rho)\left(4\pi rp +\frac{m}{r^2}-\frac{q^2}{r^3}\right)e^{2\Lambda}+\frac{q}{4\pi r^4}\frac{dq}{dr},\label{tov_equation} \quad \\
&&\frac{d\Phi}{dr}=-\frac{1}{\rho+p}\left(\frac{dp}{dr}-\frac{q}{4\pi r^4}\frac{dq}{dr}\right),\label{dg11dr}
\end{eqnarray}
where the parameters $m$ and $q$ depict the mass and charge within the radius $r$, respectively. 

Eq.~\eqref{tov_equation} represents the hydrostatic equilibrium equation, also known as the Tolman-Oppenheimer-Volkoff equation \cite{tolman,oppievolkoff}, which is a modified version from its original form to include electric charge (review, for instance, Refs.~\cite{bekenstein,raymalheirolemoszanchin,siffert,alz-poli-qbh,arbanil_malheiro2015,alz-2eos-qbh}). The above set of differential equations \eqref{eq_mass}-\eqref{dg11dr} is called as the static stellar structure equations describing a charged fluid sphere.

\begin{figure*}[ht!]
\centering
\includegraphics[width=5.748cm]{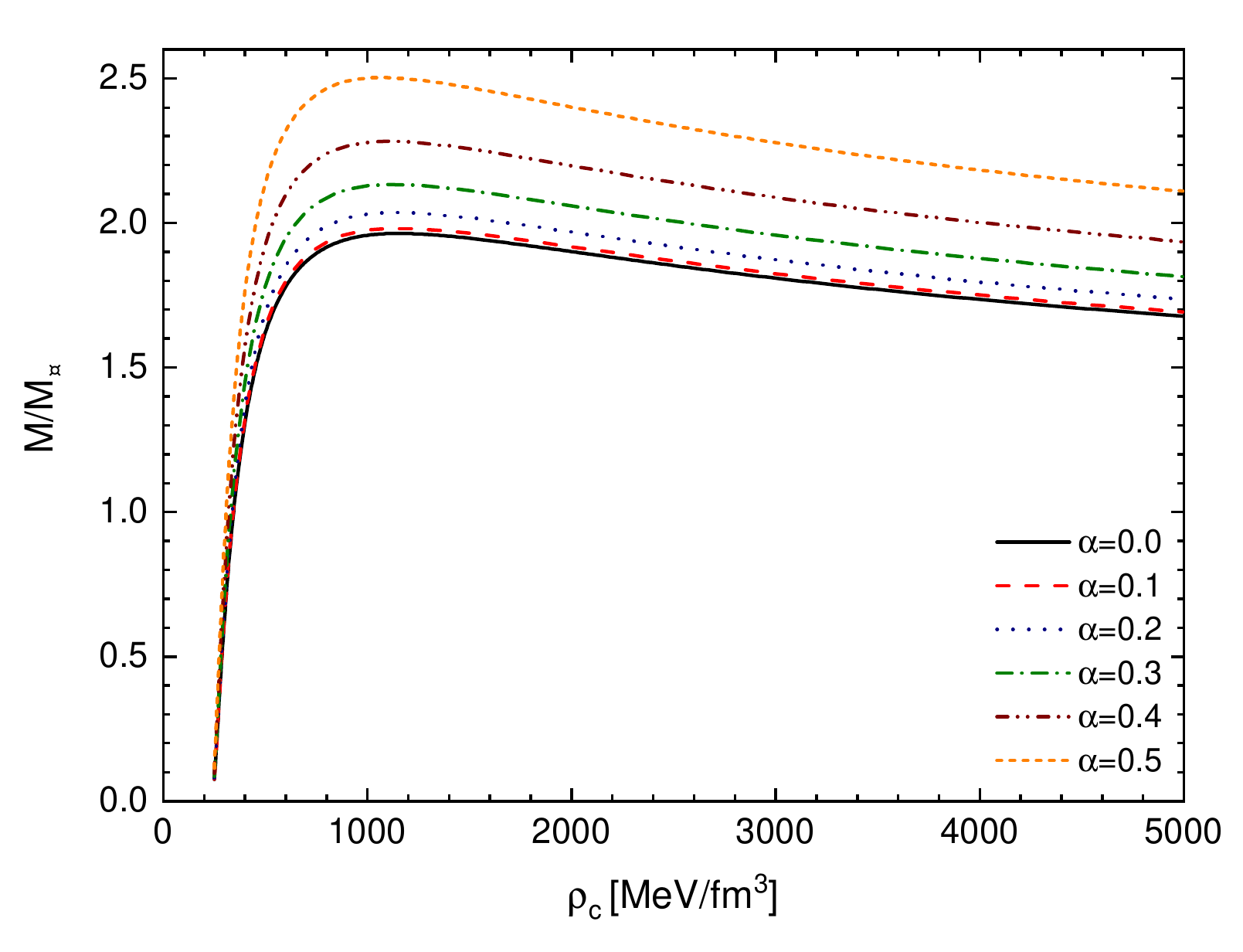} 
\includegraphics[width=5.756cm]{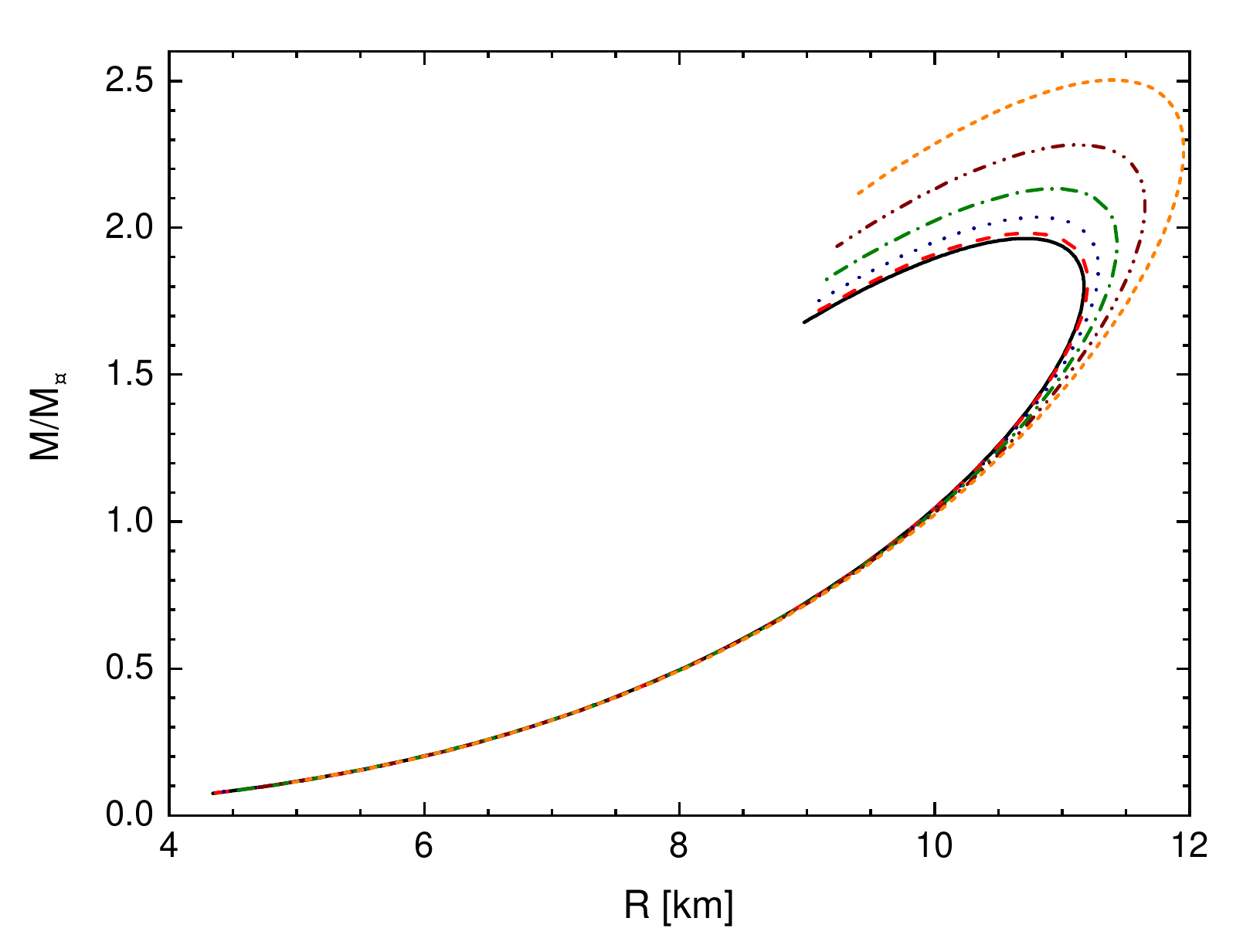}
\includegraphics[width=5.756cm]{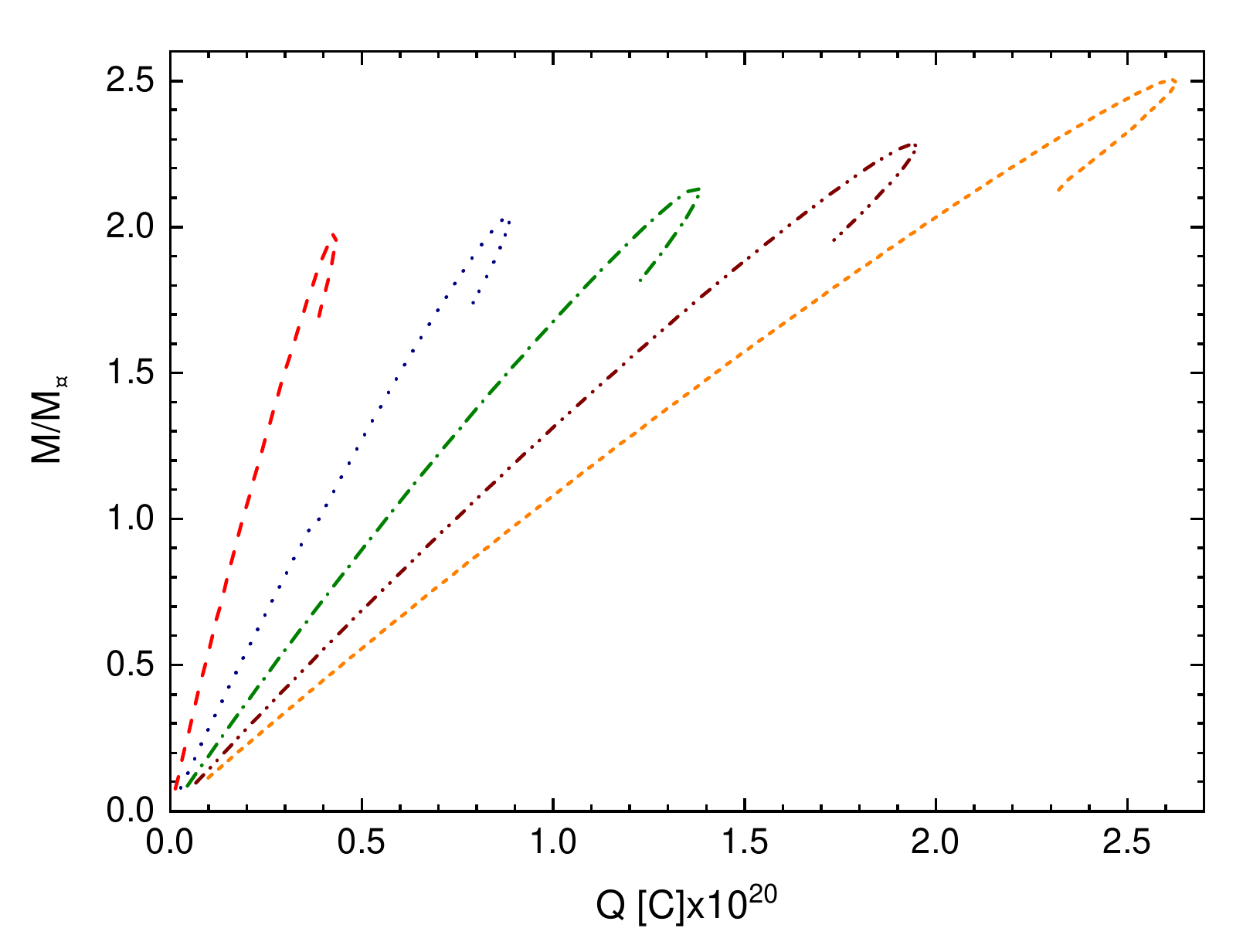}
\caption{\label{fig1} The mass in solar masses $M/M_{\odot}$ against the central energy density, radius, and total charge are plotted respectively on the left, middle, and right panels. In the three panels are employed six different values of the charge parameter $\alpha$.}
\end{figure*}

The stellar equilibrium configurations are obtained by solving Eqs.~\eqref{eq_mass}-\eqref{dg11dr}. These solutions are determined by integrating this set of equations from the center (at $r=0$) to the star's surface (at $r=R$). At the center, we use the following initial conditions
\begin{eqnarray}
&&m(0) = 0,  \quad  q(0) = 0, \quad \Lambda(0) =0,  \quad  \Phi(0) =\Phi_c,  \nonumber\\
&&\rho(0) =\rho_{c}, \quad p(0)=p_c,\quad\quad {\rm and} \quad\quad \rho_e(0)=\rho_{ec},
\end{eqnarray}
and at $r=R$ we impose that the surface is found when the pressure becomes null, i.e.,
\begin{eqnarray}
p(R)=0,
\end{eqnarray}
and, at the surface, the interior potential metrics match smoo\-thly with the exterior Reissner-Nordström solution, thus:
\begin{equation}\label{surface_condition}
e^{2\Phi}=e^{-2\Lambda}=1-\frac{2M}{R} + \frac{Q^2}{R^2},
\end{equation}
where $M$ and $Q$ represent the total mass and the total charge of the star, respectively.

\begin{figure*}[ht!]
\centering
\includegraphics[width=8.5cm]{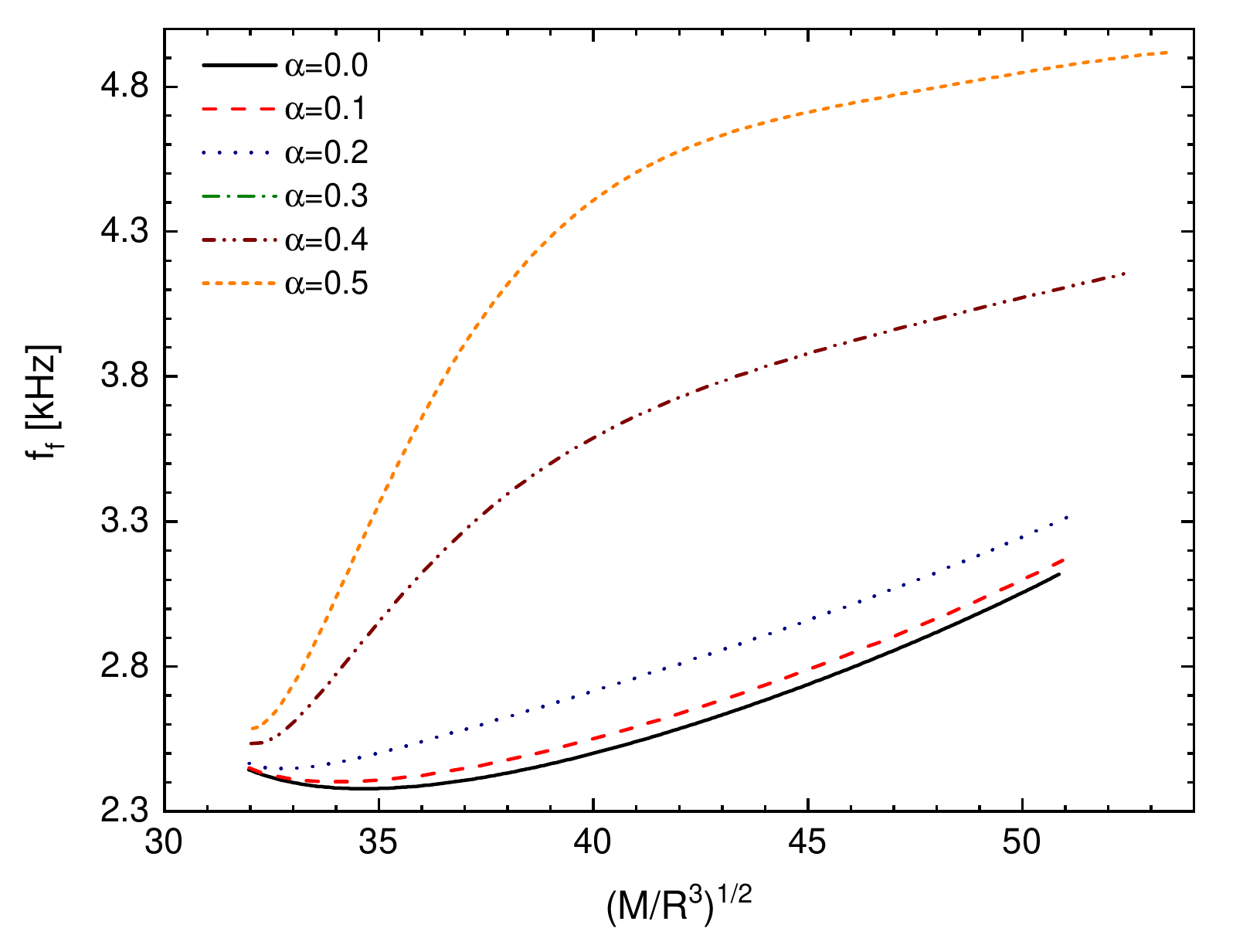} 
\includegraphics[width=8.58cm]{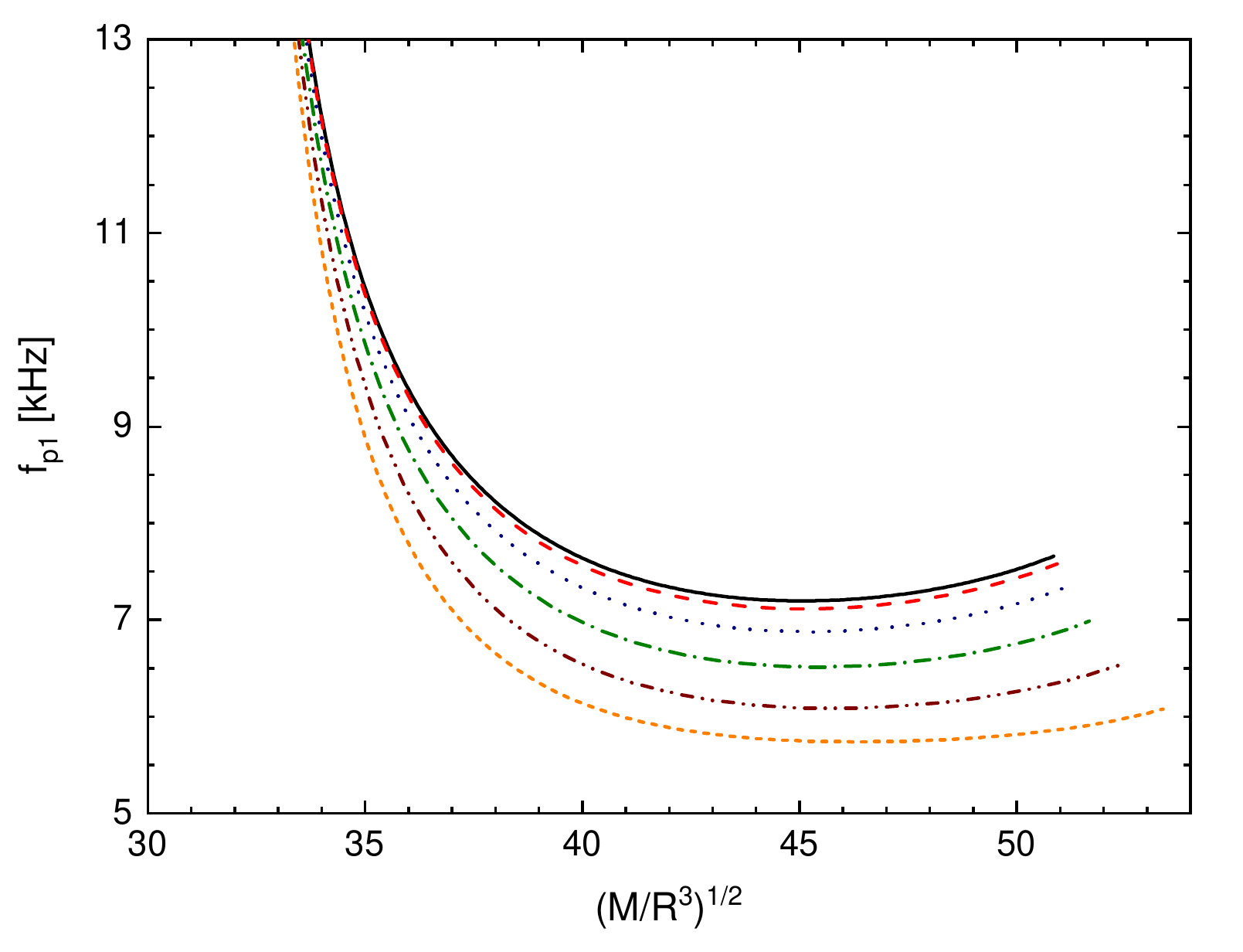}
\caption{\label{fig1.1} The $f$- and $p_1$-mode frequencies as a function of the square root of the average density $\sqrt{M/R^3}$. The results for different values of the parameter $\alpha$ are presented.}
\end{figure*}

\begin{figure*}[ht]
\centering
\includegraphics[width=8.5cm]{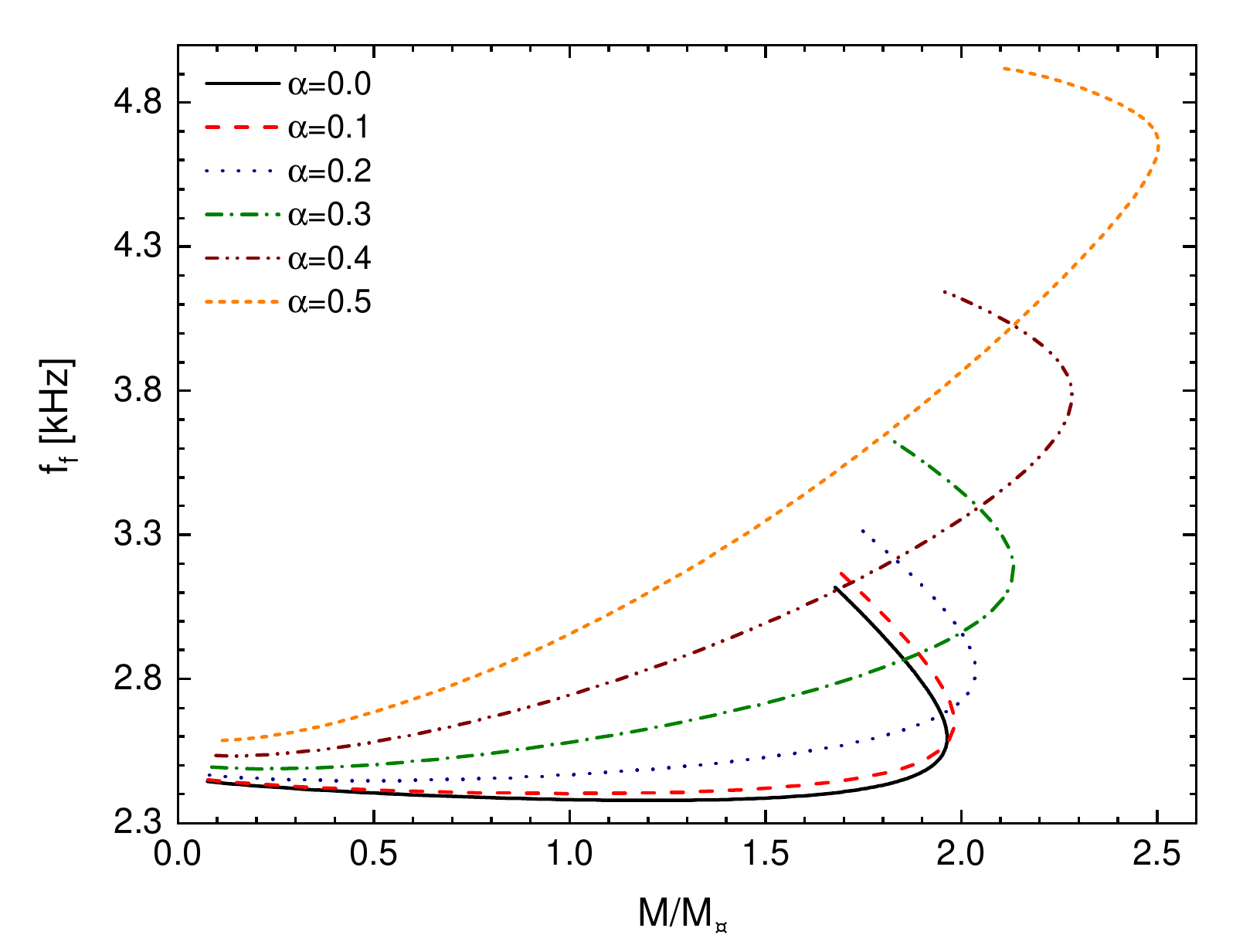} 
\includegraphics[width=8.5cm]{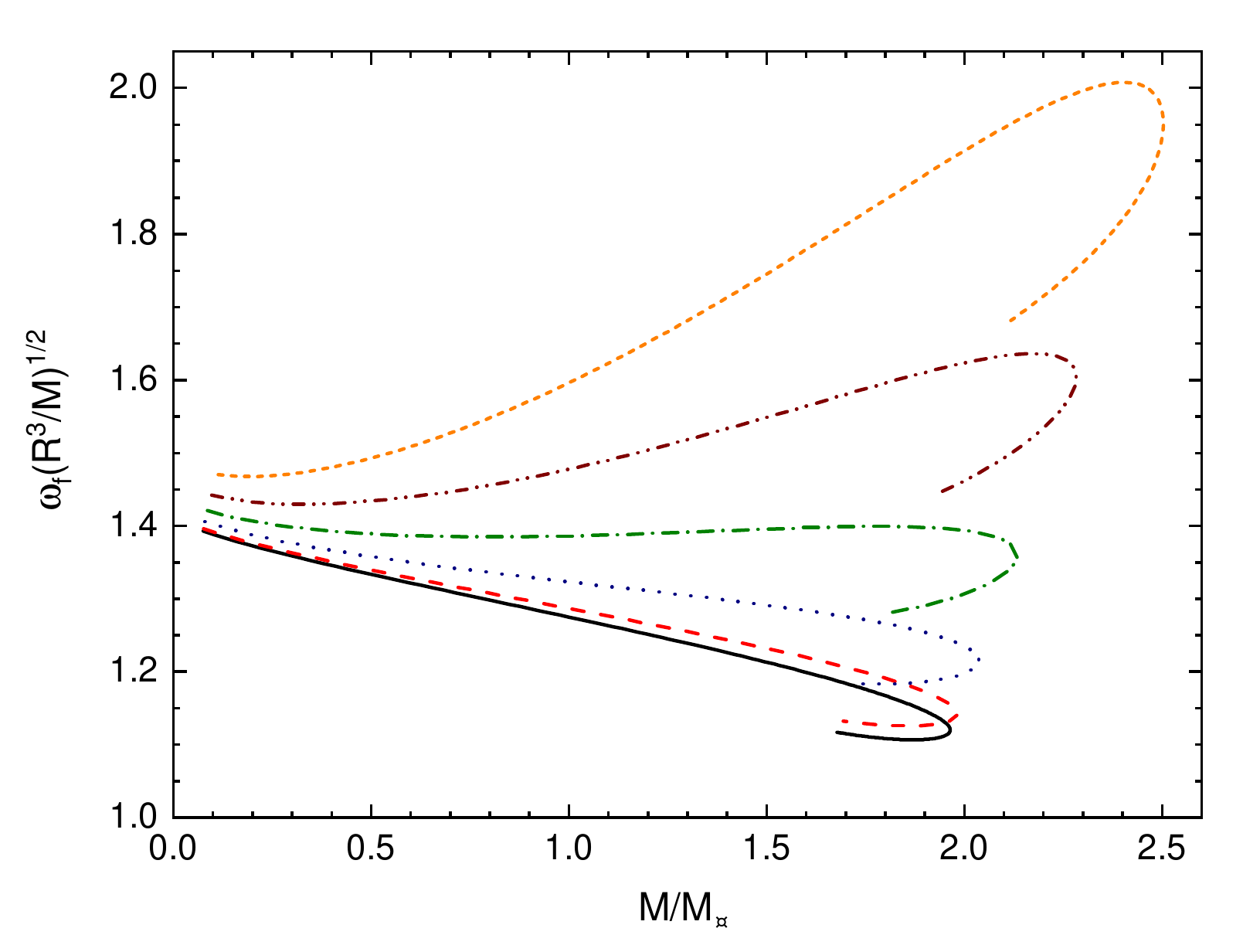} 
\includegraphics[width=8.5cm]{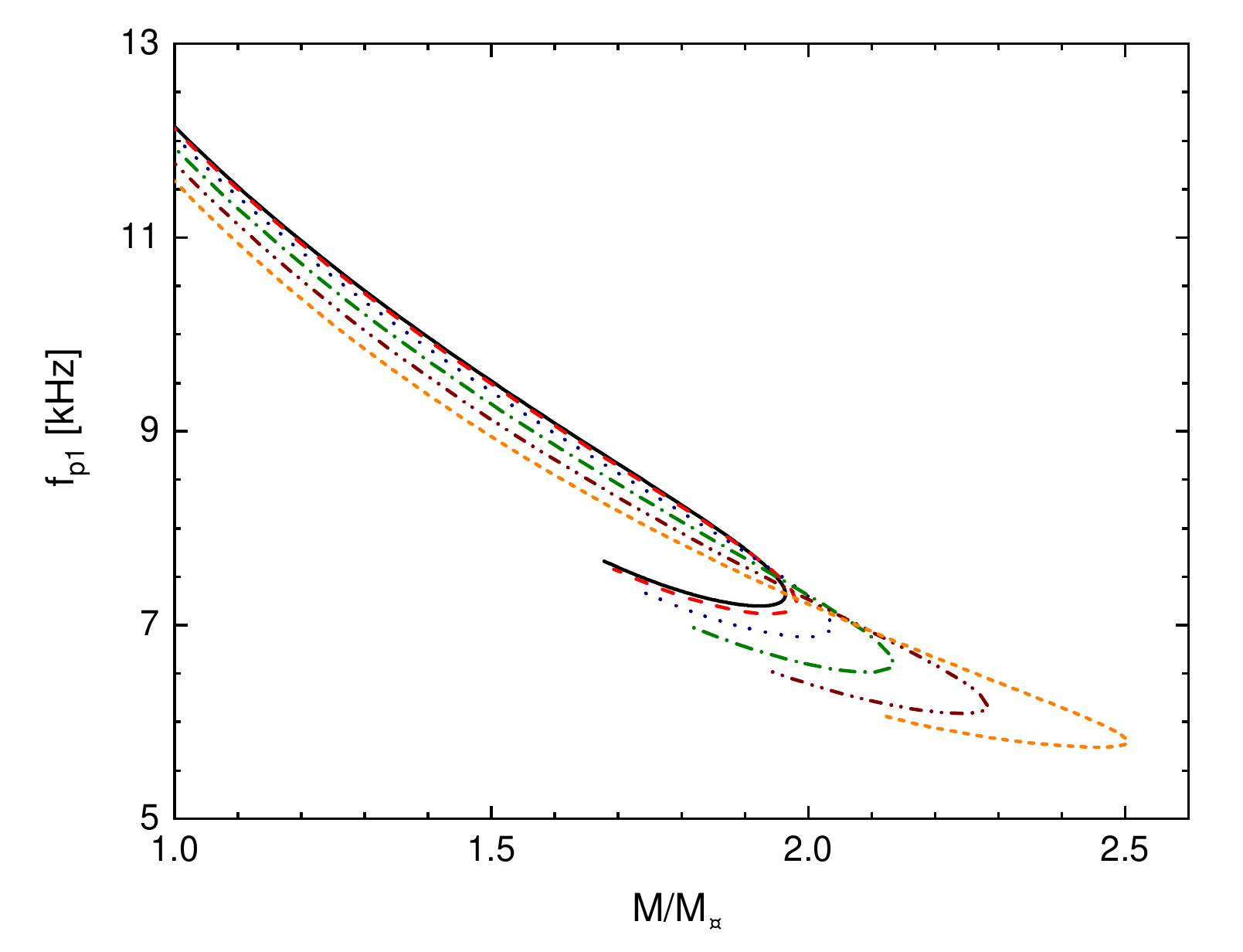} 
\includegraphics[width=8.5cm]{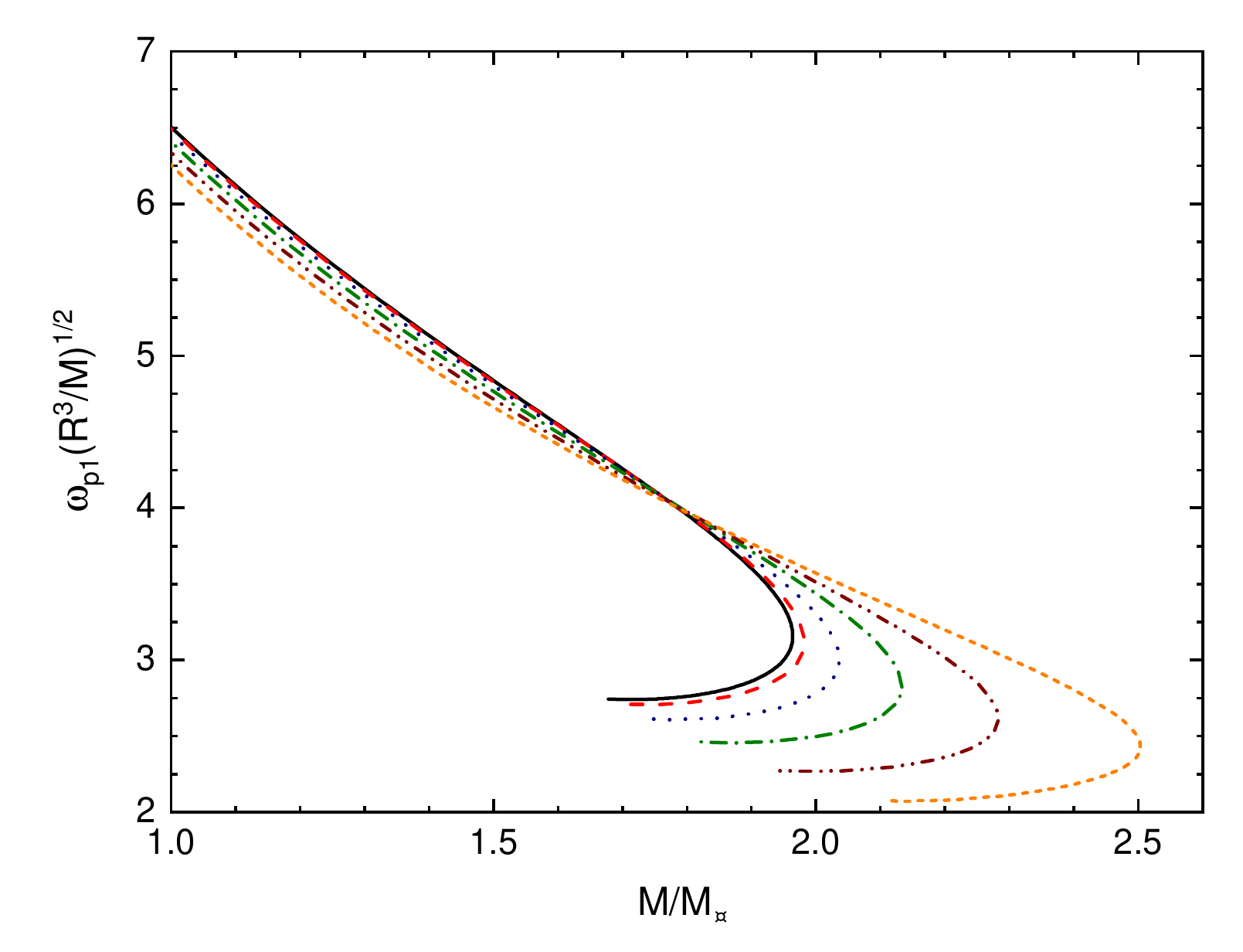} 
\caption{\label{fig2} The oscillation frequency and the normalized frequency $\omega$ as a function of the total mass $M/M_{\odot}$. On the top and bottom panels, the results for the $f$- and $p_1$-modes are presented for six values of $\alpha$.}
\end{figure*}

\subsection{Nonradial oscillations equations within the Cowling approximation}\label{subsection2}

The influence of the electric charge on the fluid pulsation modes of compact stars is investigated within the Cowling approximation, which means that the metric perturbations are null, i.e., $\delta g_{\mu\nu}=0$. 

The nonradial pulsation equations are obtained by perturbing the conservation equation of the energy-momentum tensor, namely, $\delta\left(\nabla_{\mu}T^{\mu\nu}\right)=0$. Thus, projecting it along 
the four-velocity $u_{\nu}$ and in the space orthogonal to $u_{\nu}$ by using the operator ${\cal P}_{\nu}^{\sigma} = \delta_{\nu}^{\sigma}+u^{\sigma}u_{\nu}$, we get respectively:
\begin{align}
&u^{\mu}\nabla_{\mu}\delta\rho+\delta u^{\mu}\nabla_{\mu}\left(\rho+p\right)+\left( p+\rho\right)\nabla_{\mu}\delta u^{\mu}\nonumber\\
&+(\rho+p)a_{\nu}\delta u^{\nu}+\delta F^{\nu}_{\;\;\beta}J^{\beta}u_{\nu}+F^{\nu}_{\;\;\beta}\delta J^{\beta}u_{\nu}=0,\label{nro_eq1}\\
& \ \ \nonumber \\
&\left(\delta\rho+\delta p\right)u^{\mu}\nabla_{\mu}u^{\sigma}-\left(\rho+p\right)u^{\sigma}\delta u^{\nu}u^{\mu}\nabla_{\mu} u_{\nu}+\left(g^{\mu\sigma}\right.\nonumber\\
&\left.+u^{\mu}u^{\sigma}\right)\nabla_{\mu}\delta p+\left(\rho+p\right)\delta u^{\mu}\nabla_{\mu}u^{\sigma}+\left(\rho+p\right)u^{\mu}\nabla_{\mu}\delta u^{\sigma}\nonumber\\
&-\left(\delta F_{\;\;\beta}^{\sigma}+\delta F_{\;\;\beta}^{\nu}u_{\nu}u^{\sigma}\right)J^{\beta}-\left(F^{\sigma}_{\;\;\beta}+F^{\nu}_{\;\;\beta}u_{\nu}u^{\sigma}\right)\delta J^{\beta}=0,\label{nro_eq2}
\end{align}
with $a_{\nu}=u^{\mu}\nabla_{\mu}u_{\nu}$ being the four-acceleration.

The oscillations are also described by the changing position of the fluid elements, so for this purpose, we consider that the Lagrangian displacement components take the form:
\begin{equation}\label{Eq_varsigma}
\left(\varsigma^{i}\right)=\left(e^{-\Lambda}{\tilde W},-{\tilde V}\frac{\partial}{\partial{\theta}},-\frac{{\tilde V}}{\sin^{2}\theta}\frac{\partial}{\partial{\phi}}\right)\frac{Y_{\ell}^m}{r^2},
\end{equation}
where $i$ runs from $1$ to $3$. From Eq.~\eqref{Eq_varsigma} the functions ${\tilde W}={\tilde W}(t,r)$ and ${\tilde V}={\tilde V}(t,r)$ depend of the coordinates $t$ and $r$, and $Y_{\ell}^m=Y_{\ell}^m(\theta,\phi)$ denotes the spherical harmonics. 

In this way, the perturbation of the fluid's four-velocity
\begin{equation}
(\delta u^{\mu})=\left(0,e^{-\Phi}\frac{d\varsigma^r}{dt},e^{-\Phi}\frac{d\varsigma^{\theta}}{dt},e^{-\Phi}\frac{d\varsigma^{\phi}}{dt}\right),   
\end{equation}
can be cast as:
\begin{equation}
(\delta u^{\mu})=\left(0,\frac{1}{e^{\Lambda}}\frac{\partial{\tilde W}}{\partial{t}},-\frac{\partial{\tilde V}}{\partial{t}}\frac{\partial}{\partial{\theta}},-\frac{1}{\sin^{2}\theta}\frac{\partial{\tilde V}}{\partial{t}}\frac{\partial}{\partial{\phi}}\right)\frac{Y_{\ell}^m}{r^2e^{\Phi}}.
\end{equation}

Assuming ($u^{\mu})=\left(e^{-\Phi},0,0,0\right)$, from the perturbed Max\-well's equation $\delta(\nabla_{\nu}F^{\mu\nu})=4\pi\delta J^{\mu}$, for $\mu=0$, we obtain:
\begin{equation}\label{delta_rhoe}
\delta\rho_e=-\left[\frac{d\rho_e}{dr}\frac{{\tilde W}}{r^2}+\frac{\partial{\tilde W}}{\partial r}\frac{\rho_e}{r^2}\right]\frac{Y_{\ell}^{m}}{e^{\Lambda}}-{\tilde V}\ell(\ell+1)\frac{\rho_eY_{\ell}^{m}}{r^2},
\end{equation}
and considering the relations
\begin{equation}\label{Y_X}
\frac{\delta F^{21}}{Y}=\frac{\delta F^{31}}{\partial X/\partial\theta}=-\frac{\delta F^{32}}{\partial X/\partial r}=\frac{4\pi e^{-\Phi-\Lambda}}{r^2\sin\theta},
\end{equation}
for $\mu=1,2,$ and $3$, we have respectively:
\begin{eqnarray}\label{delta_F}
(\delta F^{j0})=4\pi\rho_e\left(\frac{{\tilde W}}{e^{\Lambda}},-{\tilde V}\frac{\partial}{\partial{\theta}},-\frac{1}{\sin^2\theta}{\tilde V}\frac{\partial}{\partial{\phi}}\right)\frac{Y_{\ell}^{m}}{r^2e^{\Phi}},
\end{eqnarray}
with $j$ running from $1$ to $3$. Remark that, in Eq.~\eqref{Y_X}, the variable $Y=Y(\phi)$ is function of the coordinate $\phi$, and $X=X(r,\theta)$ depends on the coordinates $r$ and $\theta$. 

In view of Eqs.~\eqref{nro_eq1} and \eqref{delta_F}, we have that $\delta\rho$ is given by the following expression
\begin{eqnarray}
\delta\rho&=&-\frac{d\rho}{dr}\frac{{\tilde W}Y_{\ell}^{m}}{r^2e^{\Lambda}}-(p+\rho)\left(\frac{\partial {\tilde W}}{\partial r}\frac{Y_{\ell}^{m}}{r^2e^{\Lambda}}\right.\nonumber\\
&&\left.+{\tilde V}\ell(\ell+1)\frac{Y_{\ell}^{m}}{r^2}\right),
\end{eqnarray}
and considering a fluid pressure depending on the energy density of the form $p=p(\rho)$, we get
\begin{eqnarray}
\delta p&=&-\frac{dp}{dr}\frac{{\tilde W}Y_{\ell}^{m}}{r^2e^{\Lambda}}-(p+\rho)\frac{dp}{d\rho}\left(\frac{\partial {\tilde W}}{\partial r}\frac{Y_{\ell}^{m}}{r^2e^{\Lambda}}\right.\nonumber\\
&&\left.+{\tilde V}\ell(\ell+1)\frac{Y_{\ell}^{m}}{r^2}\right).
\end{eqnarray}

By using Eq.~\eqref{delta_F}, we found that the explicit form of Eq. \eqref{nro_eq2} for $\sigma=r$ and $\theta$ are, respectively:
\begin{eqnarray}
&&\frac{(p+\rho)}{r^2e^{2\Phi-\Lambda}}\frac{\partial^2{\tilde W}}{\partial t^2}Y_{\ell}^{m}+(\delta p+\delta\rho)\frac{d\Phi}{dr}+\frac{\partial\delta p}{\partial r}\nonumber\\
&&\hspace{1cm}=-\frac{4\pi\rho_e^2e^{\Lambda}{\tilde W}}{r^2}Y_{\ell}^{m}+\frac{qe^{\Lambda}\delta\rho_e}{r^2},\label{W_two_points}\\
& \ \ \nonumber \\
&&\frac{\partial\delta p}{\partial\theta}-\frac{(p+\rho)}{e^{2\Phi}}\frac{\partial^2{\tilde V}}{\partial t^2}\frac{\partial Y_{\ell}^{m}}{\partial\theta}=4\pi\rho_e^2{\tilde V}\frac{\partial Y_{\ell}^{m}}{\partial\theta}.\label{V_two_points}  \qquad
\end{eqnarray}

Now, assuming the perturbative variables harmonically time-dependent of the form ${\tilde W}(t,r)=W(r)e^{i\omega t}$ and ${\tilde V}(t,r)=V(r)e^{i\omega t}$, Eq.~\eqref{V_two_points} and Eq.~\eqref{W_two_points}-$d[$Eq.~\eqref{V_two_points}$]/dr$ respectively yield
\begin{align}
    &\frac{dW}{dr} =\frac{d\rho}{dp}\left[W\frac{d\Phi}{dr}+\frac{\omega^2r^2V}{e^{2\Phi-\Lambda}} -\frac{\rho_e e^{\Lambda}}{r^2(p+\rho)}\left(Wq\right.\right.\nonumber\\
    &\left.\left.+4\pi\rho_e r^4V\right)\right]  
    - \ell(\ell+1)e^{\Lambda}V,  \label{nro_eq_w}  \\
    & \ \ \nonumber \\
    &\frac{dV}{dr} =\frac{-e^{2\Phi}}{\omega^2(p+\rho)-4\pi\rho_e^2e^{2\Phi}} \left[W\left[-\frac{4\pi\rho_e^2e^{\Lambda}}{r^2}-\frac{q\rho_e}{r^4}\frac{d\rho}{dp}\frac{d\Phi}{dr}\right.\right.  \nonumber \\
    &\left.\left.+\omega^2(p+\rho)\frac{e^{\Lambda-2\Phi}}{r^2}-\frac{q}{r^4}\frac{d\rho_e}{dr}+\frac{q^2\rho_e^2e^{\Lambda}}{(p+\rho)r^6}\frac{d\rho}{dp}\right]\right.\nonumber\\
    &\left.+\left[\frac{4\pi q\rho_e^3e^{\Lambda}}{r^2(p+\rho)}\frac{d\rho}{dp}+\frac{\omega^2q\rho_e}{e^{2\Phi-\Lambda}r^2}-8\pi\rho_e\frac{d\rho_e}{dr}\right.\right.\nonumber\\    &\left.\left.-2\frac{d\Phi}{dr}\omega^2(p+\rho)e^{-2\Phi}-\left[1+\frac{d\rho}{dp}\right]4\pi\rho_e^2\frac{d\Phi}{dr}\right]V\right]   , \label{nro_eq_v} 
\end{align}
with $\omega$ being the oscillation eigenfrequency. 
The nonradial oscillation equations in the Cowling approximation reported in Ref.~\cite{sotani2011} are recovered considering $\rho_e=0$. It is important to say that Eqs.~\eqref{nro_eq_w} and \eqref{nro_eq_v} can also be obtained by assuming $\delta F^{21}=\delta F^{31}=\delta F^{32}=0$ in Eq.~\eqref{Y_X}. Consequently, for this model, we can indicate that fluid pulsation modes are independent of the small magnetic perturbations generated for the nonradial perturbations.

Eqs.~\eqref{nro_eq_w} and \eqref{nro_eq_v} will be numerically integrated from the center $(r=0)$ to the surface of the star $(r=R)$. At $r=0$, it is taking into account that the functions $W$ and $V$ adopt the forms:
\begin{equation}\label{WVcenter}
W=Cr^{\ell+1},\quad\quad V=-C\frac{r^{\ell}}{\ell},
\end{equation}
where $C$ represents a dimensionless constant. In addition, at $r=R$, it is found that
\begin{equation}\label{surface_condition_NRO}
W\frac{d\Phi}{dr}+\frac{\omega^2r^2V}{e^{2\Phi-\Lambda}}-\frac{\rho_e e^{\Lambda}}{r^2(p+\rho)}\left(4\pi\rho_e r^4V+Wq\right)=0.
\end{equation}

From here on, in order to compare our findings with those reported in the literature (see e.g.~Ref.~\cite{arbanil_lenzi2023}), we focus on the quadrupole moment ($\ell=2$).

\subsection{Equation of state}\label{section_eos}

For the fluid contained inside the star, we consider that the pressure and energy density are connected through the MIT bag model EoS. As is known, this equation represents a confined fluid, which is composed of up, down, and strange quarks \cite{witten1984}. In this way, in order to continue the study carried out in Ref.~\cite{arbanil_malheiro2015}, we use
\begin{equation}
    p=\frac{1}{3}\left(\rho-4\,{\cal B}\right),
\end{equation}
and we consider ${\cal B} = 60$ MeV/fm$^3$. This value is within the range of the bag constant, $57$ and 94 MeV/fm$^3$, where the theoretical hypothesis that strange matter can be the ground state of strongly interacting matter and could appear in compact stars is verified \cite{farhi1984}.

\subsection{Numerical method}

To study how electric charge affects the oscillation spectrum of strange stars---once described the equation of state and the electric charge density profile---the stellar structure equations \eqref{eq_mass}-\eqref{dg11dr} and nonradial oscillation equations \eqref{nro_eq_w}-\eqref{nro_eq_v} are solved simultaneously.

To analyze the fluid perturbation modes for some values of $\rho_{ec}$ and $\rho_c$, we first integrate Eqs.~\eqref{eq_mass}-\eqref{tov_equation} using the fourth-order Runge-Kutta method. In this way, once the radial functions $p$, $\rho$, $\rho_e$, $m$, and $\Phi$ have been obtained, we solve Eq.~\eqref{dg11dr} through the shooting method. This process starts by assuming a test value for $\Phi_c$, and whether after the iteration the condition \eqref{surface_condition} is not satisfied, the value of $\Phi_c$ is rectified to reach the desired condition. This means that we have to check if the metric potential $\Phi$ joints perfectly with the metric outside the star.

The solution of the nonradial oscillations equations is realized by integrating the set of Eqs.~\eqref{nro_eq_w}-\eqref{nro_eq_v} from the center to the surface of the compact object with the regularity conditions \eqref{WVcenter} for the functions $W$ and $V$ at the center. This procedure starts by considering the correct value of $\Phi_c$, a value of $\rho_{ec}$ and $\rho_c$, and one proof of $\omega^2$ into the set of Eqs.~\eqref{eq_mass}-\eqref{dg11dr} and \eqref{nro_eq_w}-\eqref{nro_eq_v}. If the equality \eqref{surface_condition_NRO} is not fulfilled, $\omega^2$ is corrected in the next iteration until it reaches the desired precision. Thus, for the two sets of differential equations, we have to use two guess values, $\Phi_c$ and $\omega^2$.

\section{Oscillation spectrum of the charged compact stars}\label{section3}

\subsection{Charge density profile}\label{densityprofile1}

To investigate the influence of the electric charge on the fluid pulsation modes of compact stars, the charge density profile must be defined. Thus, following \cite{liu_zhang_wen_2014,raymalheirolemoszanchin,siffert,alz-poli-qbh,alz-2eos-qbh,brillante_2014,goncalves_2022_supermassive,arbanil_malheiro2015,goncalves_2020,panotopoulos_tangphati_2022,goncalves_2022_fudamental}, we consider
\begin{equation}\label{charge_density_profile}
\rho_e=\alpha\rho,
\end{equation}
with $\alpha$ being a dimensionless constant and takes values in the range $0\leq\alpha<1$, where $\alpha= 0$ indicates the absence of charge. This relation was employed to study the effect of the electric charge on the structure of white dwarfs \cite{liu_zhang_wen_2014}, polytropic stars \cite{raymalheirolemoszanchin,siffert,alz-poli-qbh,alz-2eos-qbh}, hybrid stars 
\cite{brillante_2014,goncalves_2022_supermassive}, and strange stars \cite{arbanil_malheiro2015, goncalves_2020, panotopoulos_tangphati_2022, Zhang2021}. The electric charge influence on the radial pulsation modes of strange stars  \cite{arbanil_malheiro2015,goncalves_2020} and hybrid stars \cite{brillante_2014,goncalves_2022_fudamental} was also investigated.

Throughout this section, we will study equilibrium configurations whose total mass values are constrained by compact objects reported by observation. In this sense, the masses of the equilibrium configurations will be constrained by the event PSR J$0740+6620$, reported by the NICER team \cite{miller2021,riley2021}, whose gravitational mass is $2.08\pm0.07M_{\odot}$ and whose radius is in the interval $R=12.30^{+1.30}_{-0.98} [\rm km]$. Thus, the values of $\alpha$ that allow to fulfill such condition must be within the interval $0\leq\alpha\leq0.5$.

\subsection{Results}\label{section_results1}


In Fig.~\ref{fig1}, the total mass (given in solar masses $M_{\odot}$) as a function of the central energy density, radius, and total charge are respectively plotted on the left, middle, and right panels; where the results for different values of $\alpha$ are shown. The central energy density considered is within the interval $\rho_c \in [250, 5000]\ \rm MeV/fm^3$. In the left and right plots, in the uncharged case (i.e., when $\alpha=0$), we observe the typical behavior of $M(\rho_c)$ and $M(R)$, see e.g.~Refs.~\cite{arbanil_malheiro2015,arbanil_lenzi2023,arbanil2020,arbanil_malheiro_2016} for further details. Meanwhile, for the charged case, i.e., when $\alpha\neq0$, we obtain an increment of the total mass, radius, and total charge with $\alpha$ (see also \cite{arbanil_malheiro2015} for results with $\alpha\geq0.5$).

The $f$ and $p_1$ oscillations mode frequencies against the square root of the average density are respectively shown on the left and right-hand side of Fig. \ref{fig1.1} for six values of $\alpha$. From the figure, we note that the dependence of the $f$- and $p_1$-mode with the average density maintains its behavior for $\alpha\leq0.2$; however, for values $\alpha>0.2$, the change of the curves is more significant when the electric charge is incremented.

In Fig.~\ref{fig2} we show the gravitational wave frequency and its normalization as a function of the stellar mass. In the top panels, we display our results for the fundamental mode. In the bottom panels, we show the results for the first pressure mode, where in this case we have selected masses above approximately $1.0\, M_{\odot}$. In those plots, six different values of $\alpha$ are considered. From our outcomes, we note that for a mass interval, the increment of the dimensionless constant $\alpha$ change significantly the $f$-mode frequency value; contrary to what happens with the $p_1$-mode, where the change is not noticeable when $\alpha$ is increased.

On the other hand, by analyzing Fig.~\ref{fig2} and Table \ref{tablei}, we see that the maximum total mass $M_{\rm max}/M_{\odot}$, the oscillation modes $f$ and $p_1$  change when we increment the charge fraction $\alpha$ in $10\%$. We note that these three factors change in around $+0.90\%$, $+2.10\%$ and $-1.16\%$ when the dimensionless constant $\alpha$ goes from $0.0$ to $0.1$; these factor are altered in $+2.80\%$, $+6.70\%$ and $-3.34\%$ when $\alpha$ increases in $10\%$; they are adjusted in $+4.80\%$, $+12.0\%$ and $-5.18\%$ when $\alpha$ grows again in $10\%$; they vary in $+7.00\%$, $+19.0\%$ and $-6.71\%$ when $\alpha$  rises in $10\%$ for forth time; and finally the three factors are modified in $+9.70\%$, $+23.0\%$ and $-6.09\%$ when $\alpha$ rises in $10\%$ for fifth time. From these results, we can understand that the oscillation frequencies of the $f$-mode are more sensitive to the increase in the total charge than the total mass and the $p_1$-mode. Therefore, we can indicate that an electric charge quantity less than $10^{20}\,\rm C$ is necessary to alter the $f$ modes noticeably.


\begin{figure}[ht!]
\centering
\includegraphics[width=8.5cm]{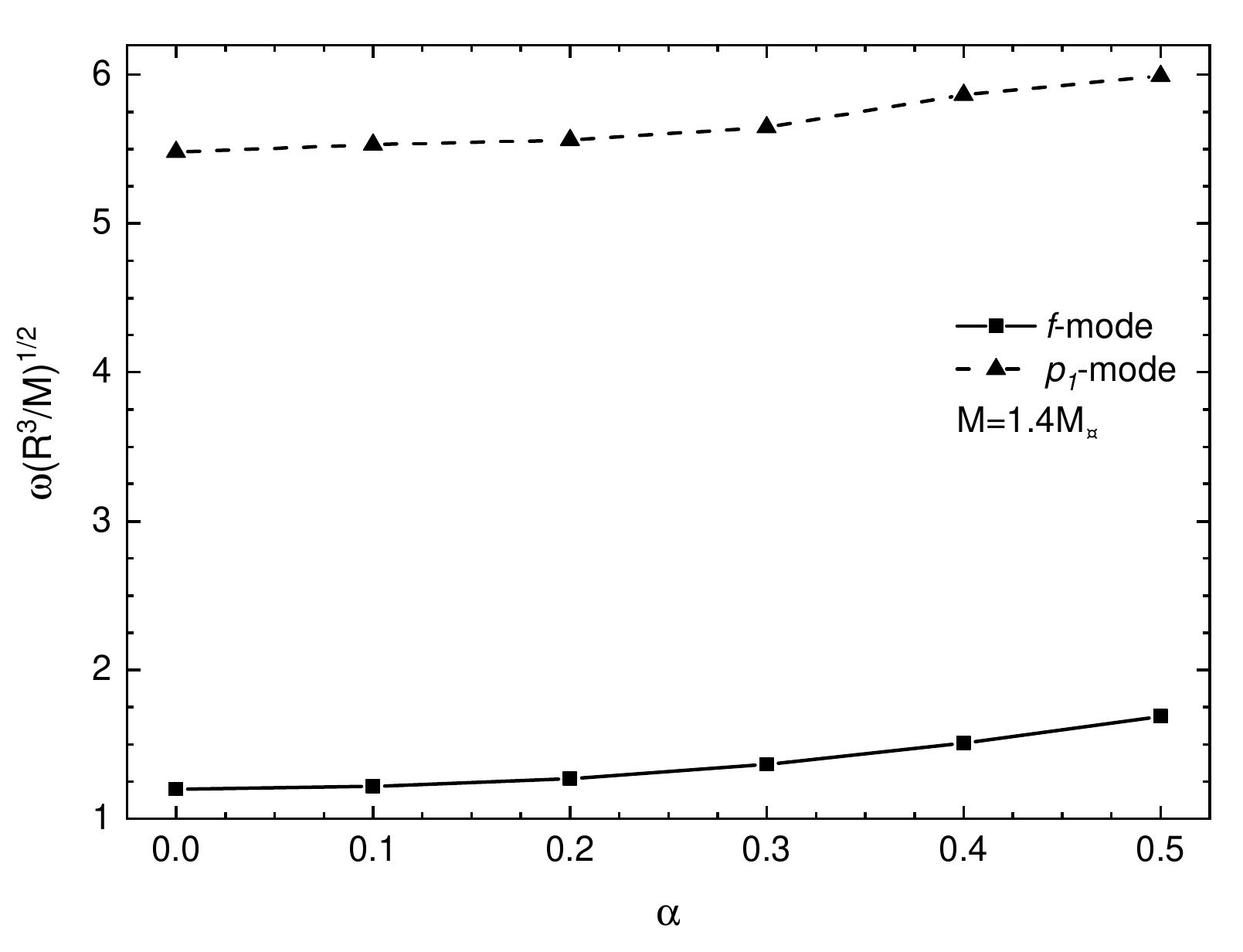} 
\caption{\label{fig3} The normalized oscillation frequency as a function of the charge parameter $\alpha$ for the total mass $M=1.4M_{\odot}$. The results for the $f$ and $p_1$ oscillation modes are presented.}
\end{figure}

In Fig.~\ref{fig3} we plot the normalized oscillation frequency against the dimensionless constant $\alpha$ for a fixed star with canonical mass $M = 1.4\, M_{\odot}$. Particularly, we can observe that the effect of the charge is to change the frequencies, for both the $f$-mode and the $p_1$-mode. Here it is important to remark that the increase in the normalized frequency is nearly linear as a function of the parameter $\alpha$.

\begin{table}[h] 
\centering
\begin{tabular}{ccccc}
\hline\hline
$\alpha$ & $M_{\rm max}/M_{\odot}$ & $Q\,[\rm C]\times10^{19}$ & $f_f\,[\rm kHz]$ & $f_{p1}\,[\rm kHz]$ \\\hline
$0.0$    & $1.9640$ & $-$ & $2.5967$ & $7.3181$ \\
$0.1$    & $1.9817$ & $4.3106$ & $2.6515$ & $7.2331$ \\
$0.2$    & $2.0363$ & $8.8222$ & $2.8292$ & $6.9912$ \\
$0.3$    & $2.1332$ & $13.743$ & $3.1797$ & $6.6189$ \\
$0.4$    & $2.2827$ & $19.388$ & $3.7813$ & $6.1751$ \\
$0.5$    & $2.5032$ & $26.169$ & $4.6524$ & $5.7991$ \\\hline\hline
\end{tabular}
\caption{\label{tablei} Equilibrium configurations with maximum total mass and their respective total charge, and $f$- and $p_1$-mode frequencies for several values of the dimensionless constant $\alpha$.}
\end{table}

\begin{figure*}[ht!]
\centering
\includegraphics[width=5.75cm]{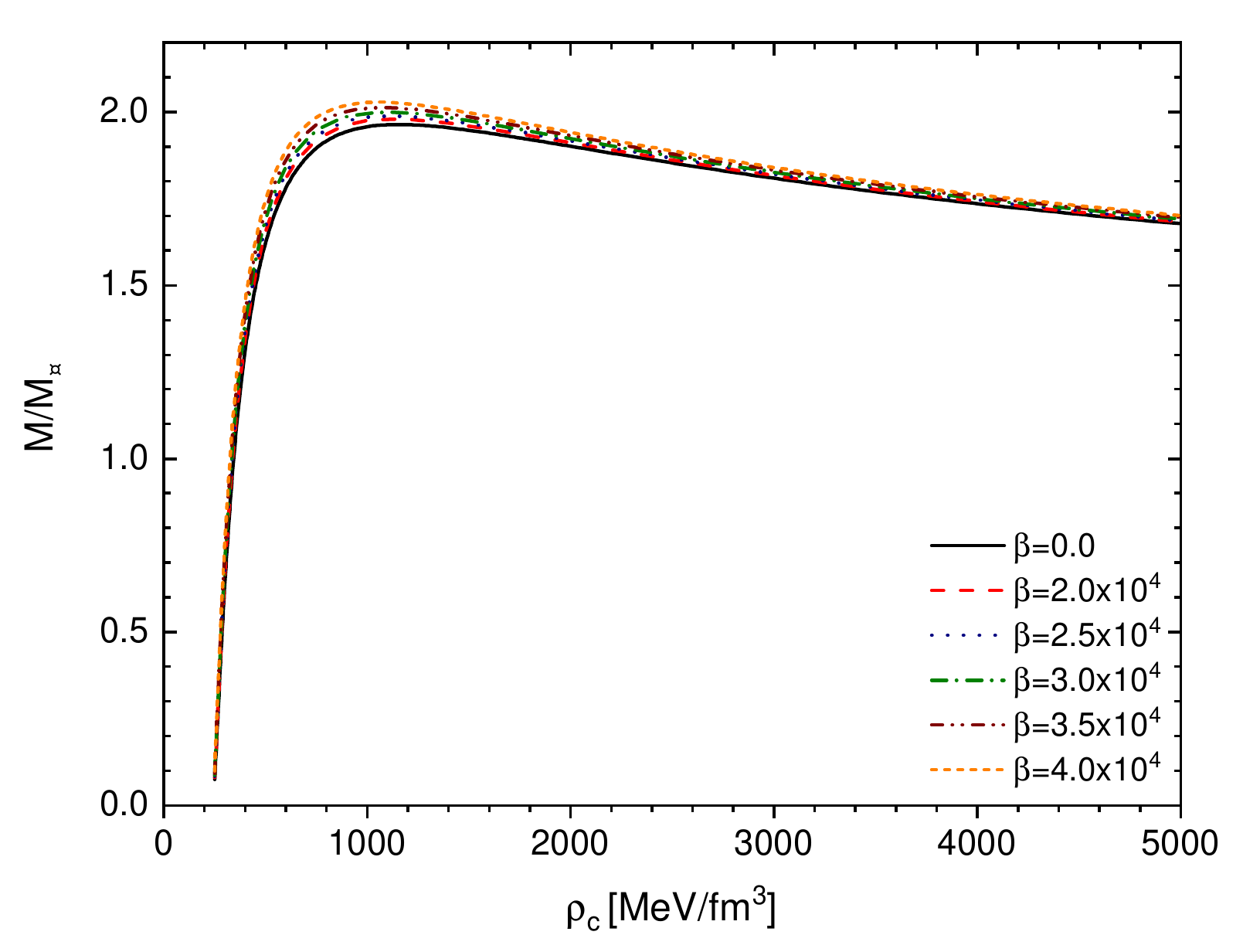} 
\includegraphics[width=5.75cm]{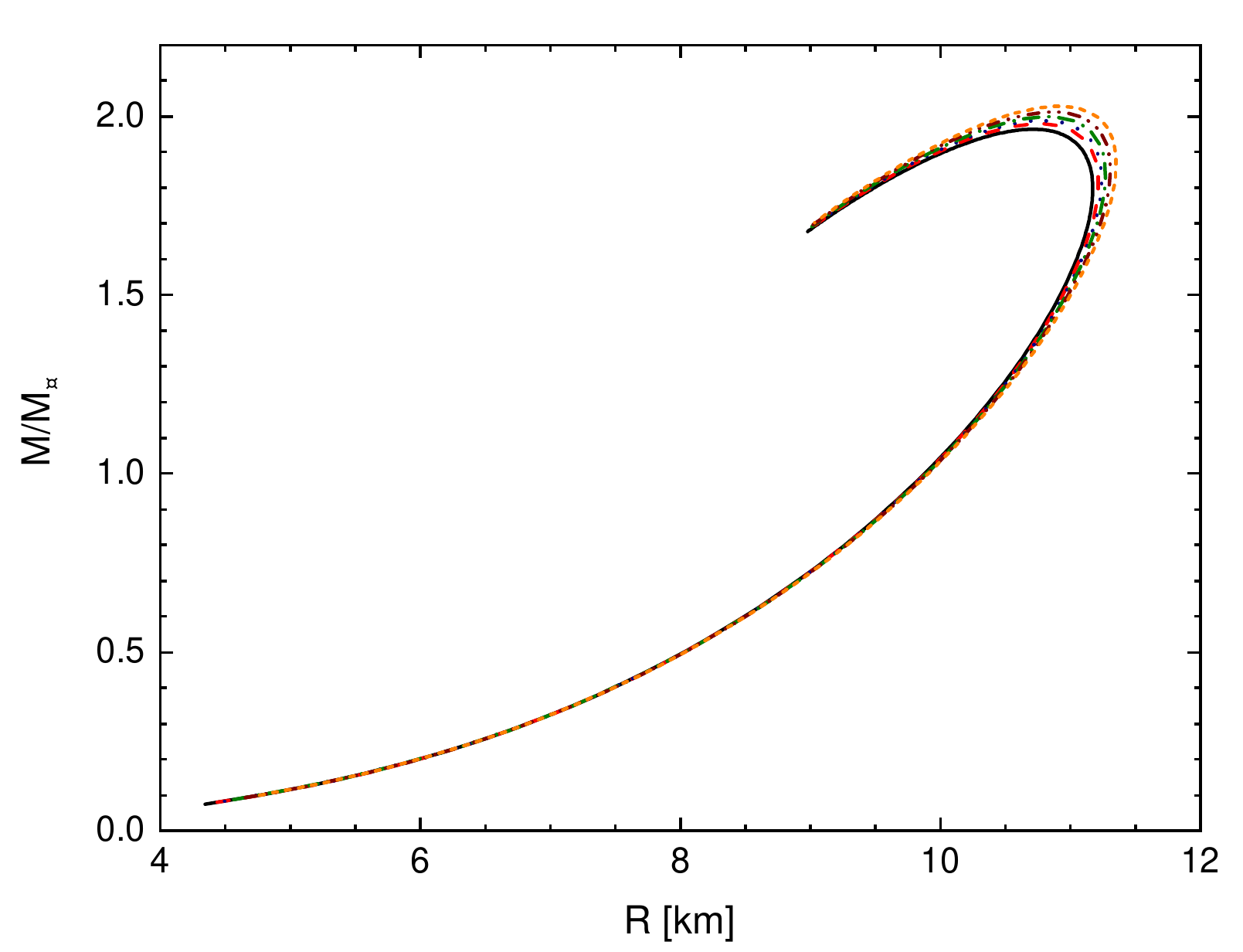}
\includegraphics[width=5.75cm]{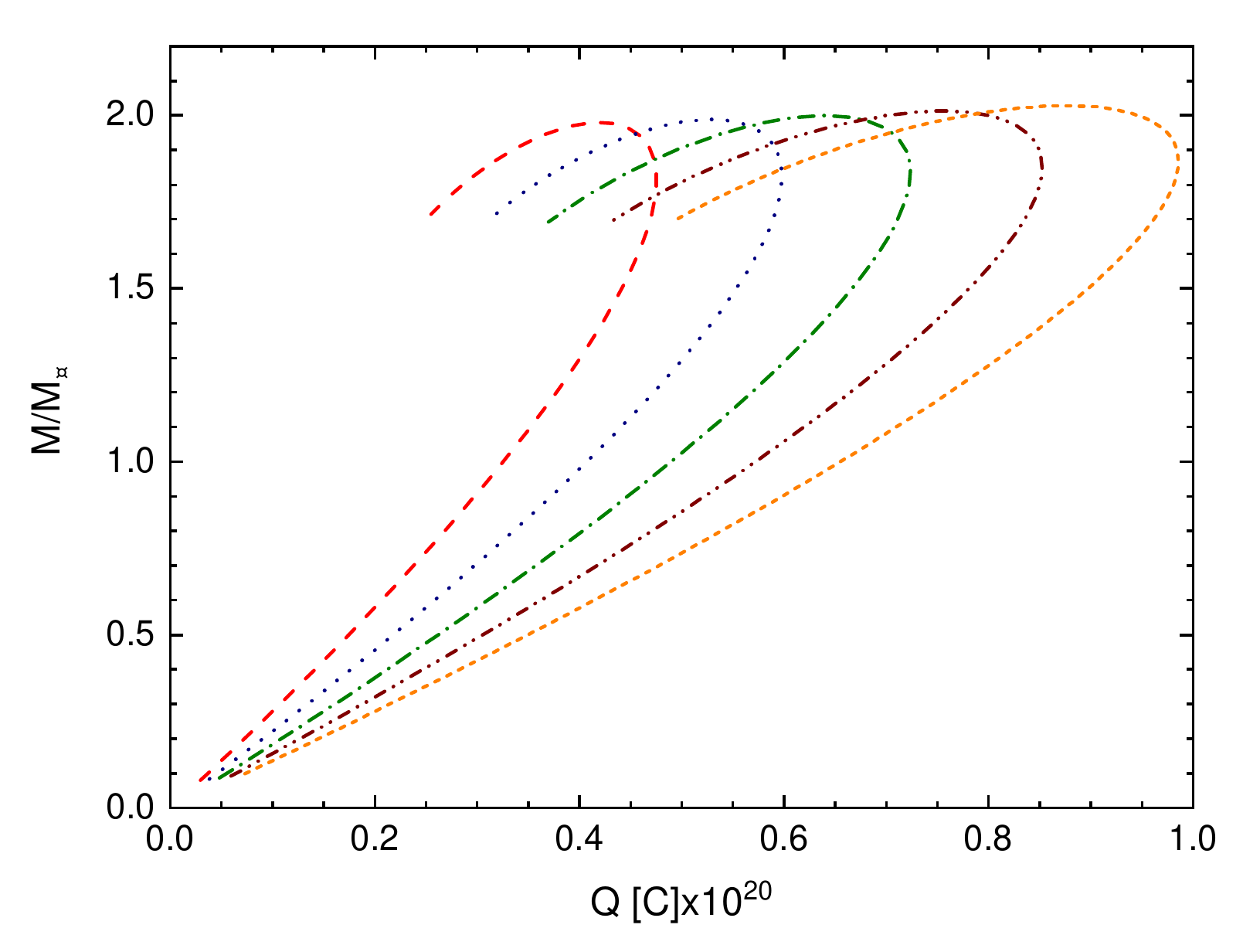}
\caption{\label{fig4} The total mass in Sun masses as a function of the central energy density, radius, and total charge are respectively presented on the left, middle, and right panels. In all plots, six different values of $\beta$ are used.}
\end{figure*}

\section{Oscillation spectrum for a different charge distribution}\label{section4}

\subsection{Charge density profile}\label{section_profile2}

To investigate the dependence of the fluid pulsation modes on the electric charge profile, we also consider the electric charge distribution of the form:
\begin{equation}
    q=Q\left(\frac{r}{R}\right)^3\equiv\beta r^3,
\end{equation}
with $\beta$ being a constant that is related to both the total charge and radius of the form $\beta=Q/R^3$. The simple case of this relation was examined in Refs.~\cite{defelice1995,defelice1999}.

\subsection{Results}

The total mass versus central energy density, radius, and total charge are presented on the left, middle, and right panels of Fig.~\ref{fig4} for six different values of $\beta$. Note that for all values of $\beta$ employed, the total mass and radius undergo an insignificant change and, in turn, the total charge reaches a maximum value of $\sim10^{20}[\rm C]$. Moreover, according to the third plot, the behavior of the $M-Q$ curves differs appreciably from those obtained for the other charge profile \eqref{charge_density_profile}.

\begin{figure*}[ht!]
\centering
\includegraphics[width=8.5cm]{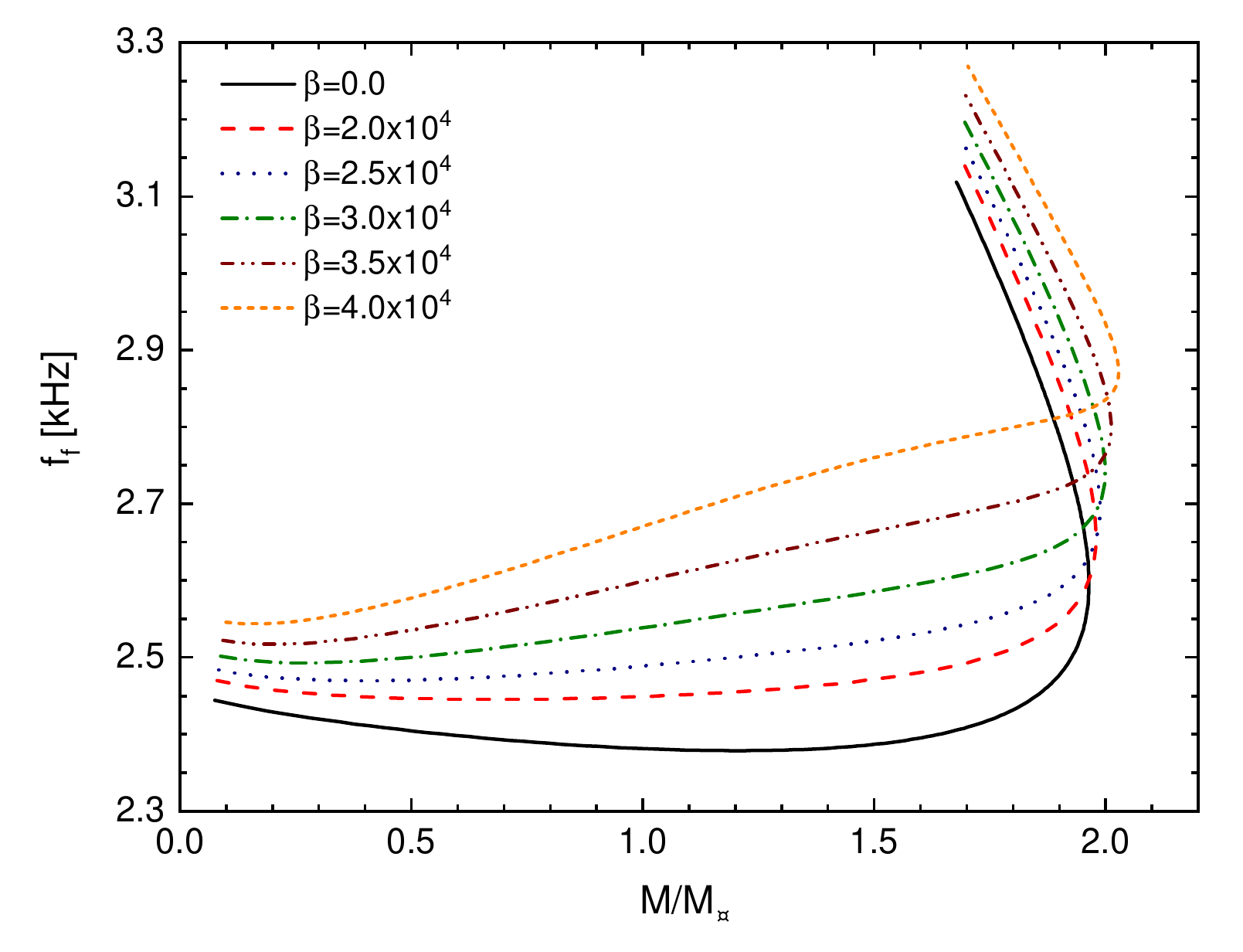} 
\includegraphics[width=8.5cm]{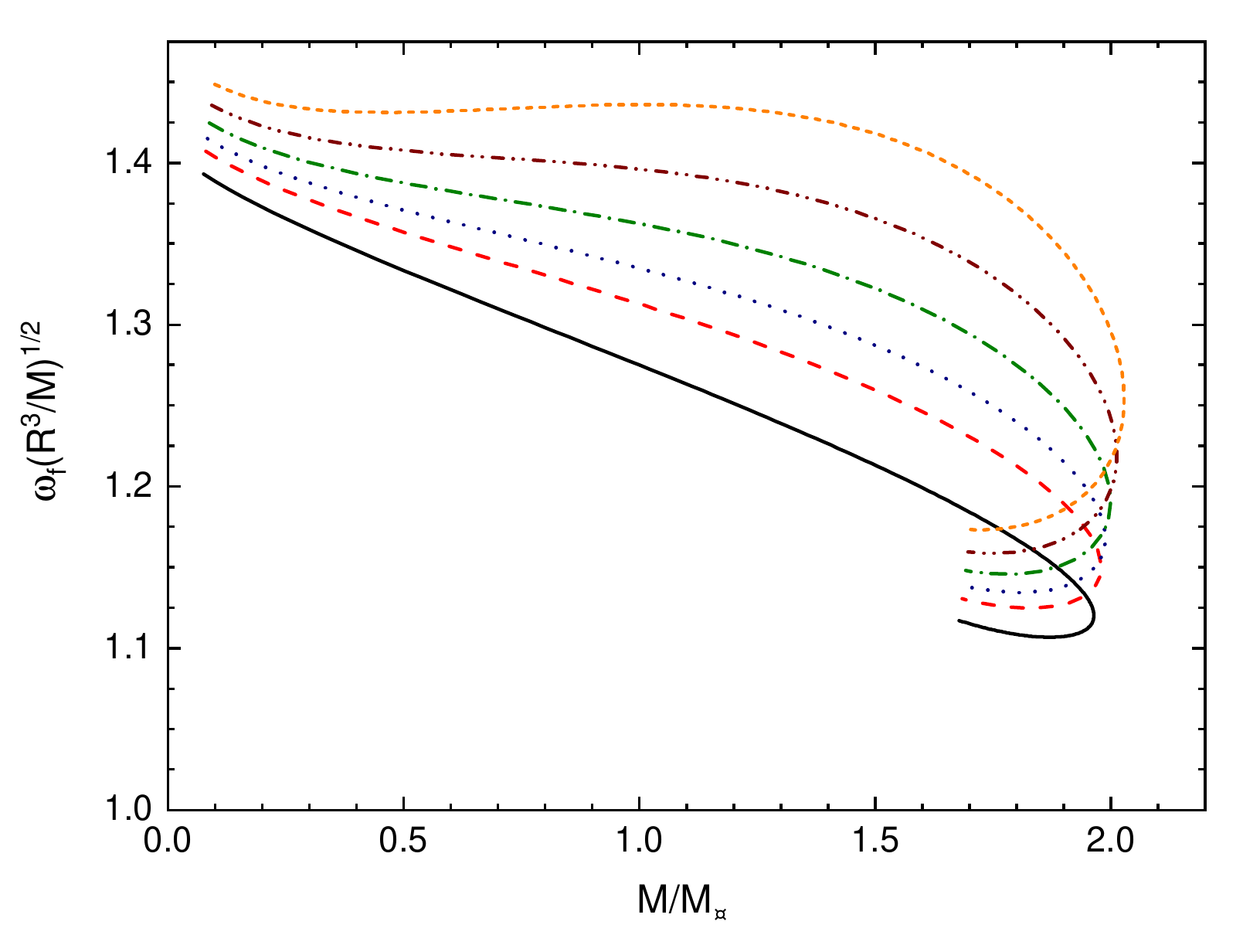} 
\includegraphics[width=8.5cm]{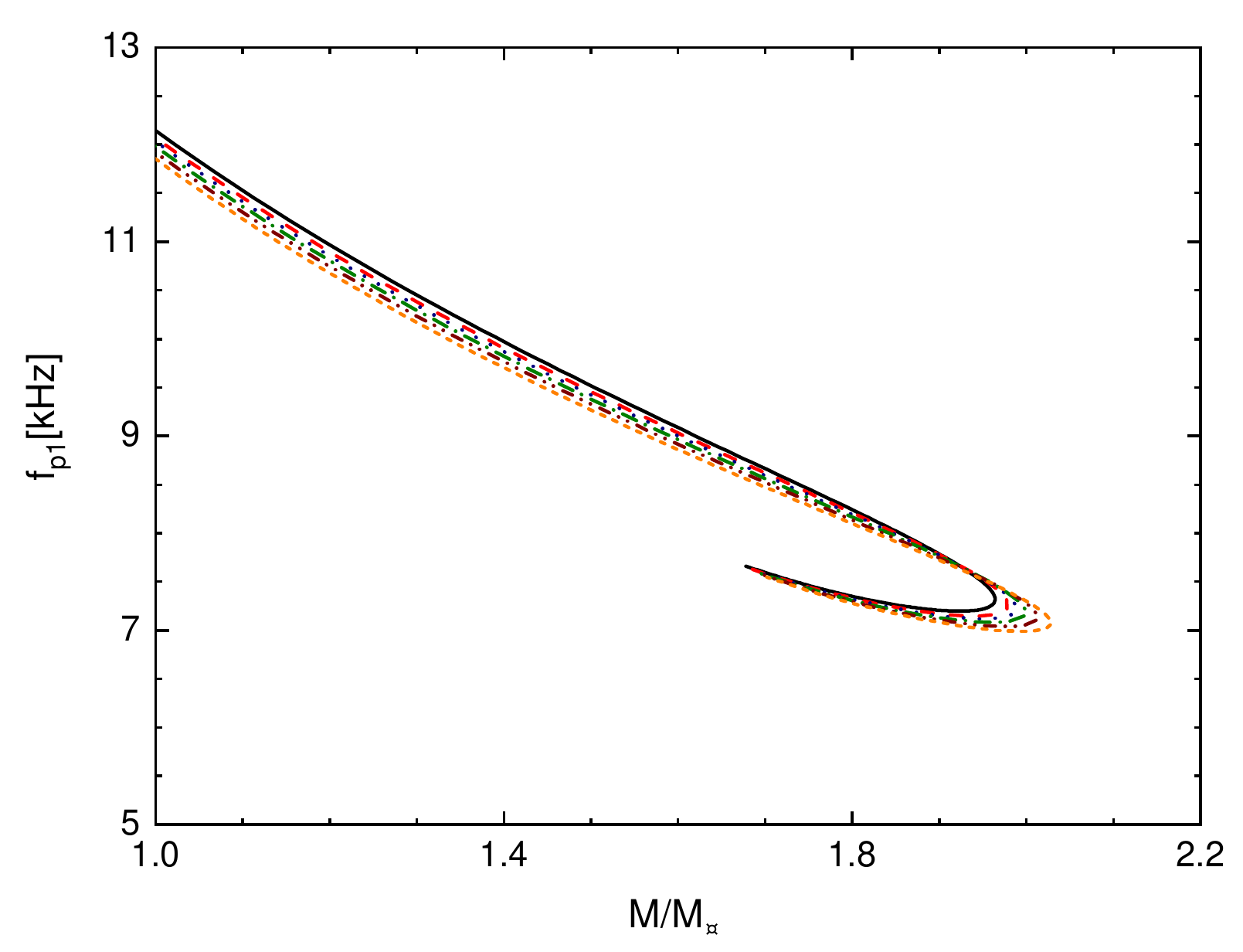} 
\includegraphics[width=8.5cm]{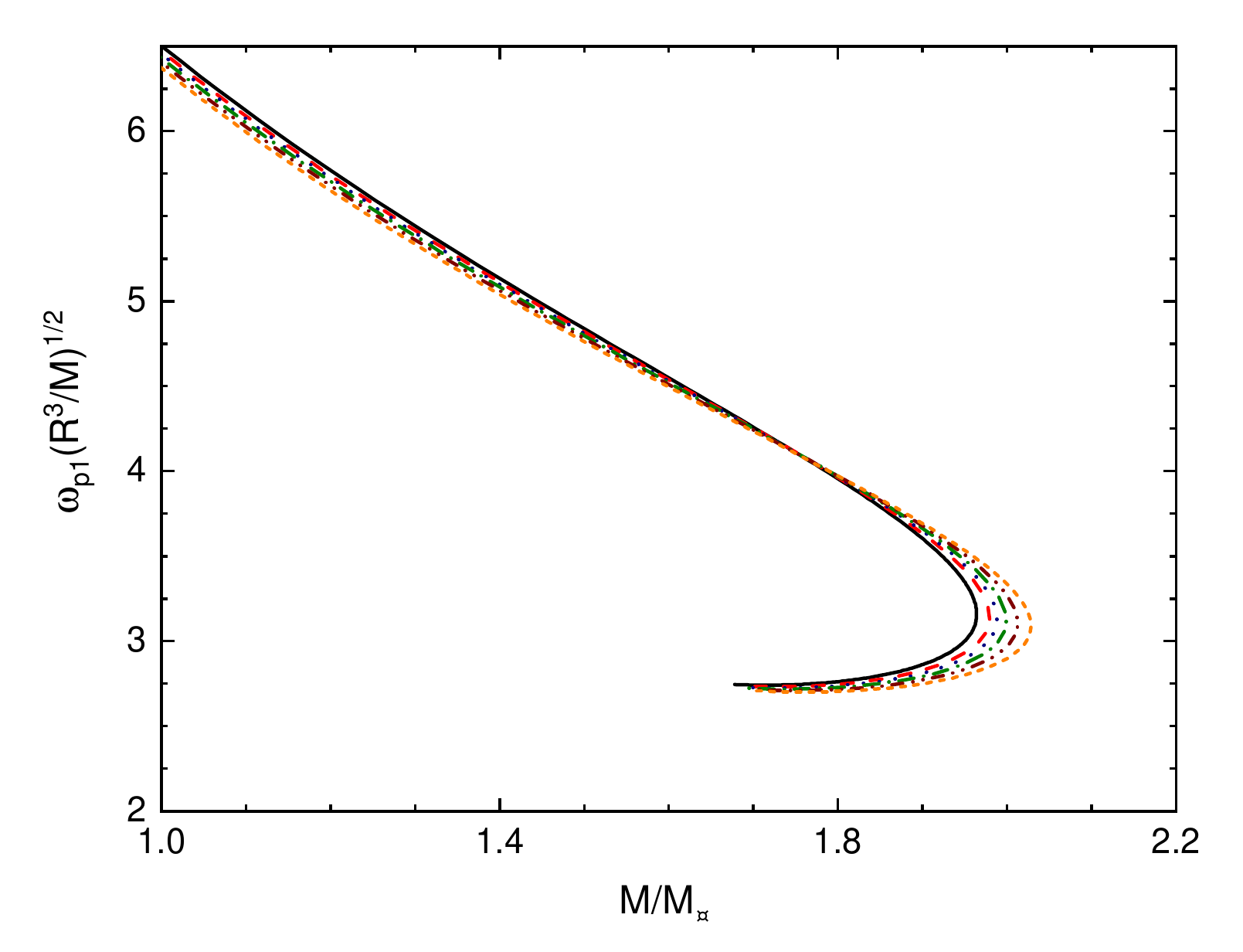} 
\caption{\label{fig5} The frequency of oscillation and the normalized frequency $\omega$ against the total mass $M/M_{\odot}$. On the top and bottom panels are respectively shown the results for the $f$- and $p_1$-modes for six values of $\beta$.}
\end{figure*}

Fig.~\ref{fig5} presents the fluid pulsation modes and their normalized frequency $\omega$ versus the total stellar mass. On the top and bottom panels are shown the results for the fundamental mode and first pressure mode, respectively. The results for the $p_1$ mode case are limited for total masses above $1.0\, M_{\odot}$. In all panels of this figure, six values of $\beta$ are considered. From the results, we can note that for some mass intervals, the $f$-mode increases significantly with the increment of $\beta$; however, the change of the $p_1$-mode with $\beta$ is imperceptible.

\section{Conclusions}\label{section5}

In this work, we have investigated the influence of the electric charge on the $f$- and $p_1$-mode of oscillation of compact stars composed of electrically charged matter. For the fluid matter, we used the MIT bag model EoS, and for the charge profile we considered a direct relation between the charge density and energy density of the form $\rho_e=\alpha\rho$; being $\alpha$ a dimensionless constant. The pulsation frequency was analyzed in terms of some values of $\rho_c$ and $\alpha$.


For this purpose, we have obtained the system of equations corresponding to the non-radial oscillations of the char\-ged star. In this sense, we have six coupled equations: four equations for the stellar structure of the charged fluid and two equations for the fluid oscillations. After that, we proceeded to find the stellar mass, total charge, and frequencies of oscillations for the compact star. To accomplish our objectives we used a standard shooting method, already described in the article. We studied the fundamental and first pressure modes, focusing on quadrupolar radiation i.e. $l = 2$. 

As our main result, we have found that the fluid oscillation modes are affected by the increment of the dimensionless constant $\alpha$. For some mass interval, we noted that the $f$-mode changes are more noticeable than the $p_1$-mode when $\alpha$ is incremented. Analyzing the equilibrium configurations with maximum total masses, we showed that the necessary charge quantity to modify the $f$-mode is $\lesssim10^{20}\, [\rm C]$. Finally, employing an electric charge distribution proportional to the radial coordinate, of the form $q=\beta r^3$, we found that the $f$-mode change is more sensitive than the $p_1$-mode when electric charge increases. Moreover, that necessary electric charge to affect the $f$-mode is less than $10^{20}\,[\rm C]$.

The nonradial oscillation of charged compact stars by using the fully general relativistic and the inverse Cowling approximation should also be investigated, we shall disseminate these studies in future works.


\begin{acknowledgements}
JDVA would like to thank Universidad Privada del Norte and Universidad Nacional Mayor de San Marcos for the financial support - RR Nº$\,005753$-$2021$-R$/$UNMSM under the project number B$21131781$. JMZP acknowledges support from ``Funda\c{c}\~ao Carlos Chagas Filho de Amparo \`a Pesquisa do Estado do Rio de Janeiro'' -- FAPERJ, Process SEI-260003/000308/2024. COVF also makes his acknowledgments for the financial support of the productivity program of the Conselho Nacional de Desenvolvimento Cient\'ifico e Tecnol\'ogico (CNPq), with Project No. $304569/2022-4$. Finally, C.H.L. would like to thank CNPq for the financial support with Projects No $401565/2023-8$ (Universal CNPq) and No $305327/2023-2$ (productivity program).
\end{acknowledgements}


\end{document}